\documentclass[preprint]{aastex}

\newcommand\cs{c_s}
\newcommand\vA{v_{\rm A}}
\newcommand\kx{{k_x}}
\newcommand\ky{{k_y}}

\newcommand\Msun{{\rm\,M_\odot}}

\newcommand\pc{{\rm\,pc}}
\newcommand\kpc{{\rm kpc}}
\newcommand\Alf{Alfv\'en }
\newcommand\torb{t_{\rm orb} }
\newcommand\Qsp{Q_{\rm sp} }
\newcommand\Qc{Q_{0,c} }
\newcommand\Smax{\Sigma_{\rm max} }
\newcommand\Kymax{K_{y,\rm max} }
\newcommand\ljsp{\lambda_{\rm J,sp} }
\newcommand\kjsp{k_{\rm J,sp} }
\newcommand\dvx{\delta u}
\newcommand\dvy{\delta v}

\newcommand\simgt{\lower.5ex\hbox{$\; \buildrel > \over \sim \;$}}
\newcommand\simlt{\lower.5ex\hbox{$\; \buildrel < \over \sim \;$}}

\shortauthors{Kim \& Ostriker}
\shorttitle{Formation and Fragmentation of Spurs}

\begin{document}

\pagebreak
\title{Formation and Fragmentation of Gaseous Spurs in Spiral Galaxies}

\author{Woong-Tae Kim and Eve C. Ostriker}
\affil{Department of Astronomy, University of Maryland \\
College Park, MD 20742-2421}

\email{kimwt@astro.umd.edu, ostriker@astro.umd.edu}

\slugcomment{Accepted for publication in the Astrophysical Journal
\hspace{1.3cm}}

\begin{abstract}
Intermediate-scale spurs are common in spiral galaxies, but perhaps most
distinctively evident in a recent image showing a quasi-regular
series of dust lanes projecting from the arms of M51 \citep{scorec01}.
We investigate, using time-dependent numerical MHD simulations, how
such spurs could form (and subsequently fragment) from 
the interaction of a gaseous interstellar medium with
a stellar spiral arm. We model the gaseous medium as a self-gravitating,
magnetized, differentially-rotating, razor-thin disk. The basic
flow shocks and compresses as it passes through a local segment of
a tightly-wound, trailing stellar spiral arm, modeled as a
rigidly-rotating gravitational potential.
We first construct one-dimensional profiles for flows with spiral shocks.
When the post-shock Toomre parameter $\Qsp$ is sufficiently small, 
self-gravity is too large for one-dimensional steady solutions to exist.
The critical 
values of $\Qsp$ are $\sim$0.8, 0.5, and 0.4 for our models with zero, 
sub-equipartition, and equipartition magnetic fields, respectively.
We then study the growth of self-gravitating perturbations in fully
two-dimensional flows, and find that spur-like structures rapidly
emerge in our magnetized models. We associate this gravitational 
instability with the magneto-Jeans mechanism, in which magnetic
tension forces
oppose the Coriolis forces that would otherwise prevent the 
coalescence of matter along spiral arms.
The shearing and expanding velocity field
shapes the condensed material into spurs as it flows downstream
from the arms.
Although we find swing amplification can help form spurs when 
the arm-interarm contrast is moderate, unmagnetized systems 
that are quasi-axisymmetrically stable are generally also stable to 
nonaxisymmetric perturbations, suggesting that magnetic effects are essential.
In nonlinear stages of evolution, the spurs in our models undergo 
fragmentation to form $\sim4\times 10^6 \Msun$ clumps, which we suggest could
evolve into bright arm and interarm \ion{H}{2} regions as seen 
in spiral galaxies.

\end{abstract}

\keywords{galaxies: ISM --- galaxies: kinematics and dynamics
--- galaxies: spiral --- galaxies: structure --- instabilities 
--- ISM: kinematics and dynamics
--- ISM: magnetic fields --- MHD --- stars: formation}

\section{Introduction}

Even the grandest of grand-design spiral galaxies abound with
substructure overlying and interlaced with the dominant twin-armed
global pattern. This substructure may take the form of feathering
between the main arms, or of spurs or larger branches jutting
nearly perpendicularly from the primary arms, then sweeping
back in the same sense at larger radii
(e.g., \citealt{lyn70,wea70,elm80,rob90}).
Secondary structures often appear in combination (e.g., \citealt{pid73}),
and are sometimes associated with chains of interarm \ion{H}{2} regions
\citep{van76,elm79}.
These intermediate-scale features occur in multiple-armed spirals as 
well as grand design types, but blend more unobtrusively into the 
complex overall structure.
Even at quite small (circumnuclear) scales, special
image processing techniques can reveal regular substructure
in spiral arms \citep{lou01}.
From morphological evidence in multiple colors combined with kinematic 
arguments, \citet{elm80} concluded that many spurs represent long-lived
wavelike phenomena.

Although spurs have long been recognized as characteristic 
features from ground-based studies, the higher resolution afforded 
by space-based platforms opens possibilities for observing these 
structures in exquisite and unprecedented detail.
In particular, the recently-released Hubble Heritage image of
the Whirlpool Galaxy M51 \citep{scorec01,sco01}
stunningly reveals that spurs project, at quasi-regular
intervals, from essentially the whole length of the two
main arms. These spurs are defined by narrow dust lanes, and
are dotted with \ion{H}{2} regions both near the main
arms and well into the interarm regions.

What is the origin of all these feathers, spurs, and branches?
Do they represent a stochastic, localized phenomenon, or does the 
global spiral pattern play a role in their formation?
Small-scale, irregularly-oriented and -separated feathering could
in principle arise in many ways. 
Proposals  include
gravitational induction of quasi-steady trailing responses to
orbiting local mass condensations \citep{jul66,byr83,byr84},
transient swing amplification of ``shearing bits and pieces'' from
a near-uniform interarm background \citep{gol65,jul66,too81},
shocking of the interstellar medium (ISM) produced by ballistic 
clouds traversing spiral arms \citep{pik70,kap74},
and percolation in a shearing environment of self-propagating
star formation (e.g., \citealt{ger78,sch86}).
Large-scale interarm branches that lie in relative radial 
isolation could be associated with ultraharmonic
resonances of the primary pattern's potential 
\citep{shu73,ees89,elm90}.

In addition to structural features for which the large-scale
spiral pattern is irrelevant, and those for which it is crucial
(via a resonance), 
there may be features that are organized in a spatially-coherent 
plan by the spiral arms, yet which develop similarly to local,
self-gravitating, shearing wavelets. We believe that the
spurs recently observed in
M51 (and likely to show themselves in other galaxies
upon high-resolution examination) represent this sort of 
hybrid phenomenon.

The idea that spiral arms may trigger the growth of smaller-scale,
self-gravitating structures -- particularly in the gaseous medium --
is long established, and supported by both observational
and theoretical studies. Observational evidence includes
the consistency with characteristic Jeans-unstable values 
of observed masses and separations of \ion{H}{1} superclouds
\citep{elm83, elm87b, kna93} and 
giant molecular associations
\citep{vog88,ran93, sak96, tho97a, tho97b, sak99}. 
In addition, \ion{H}{2} regions and star complexes are
often distributed along arms in a ``beads on a string''
pattern, with a separation of 
roughly 3 times ($\sim 1-4$ kpc) the full arm thickness,
independent of pitch angle and galactocentric radius \citep{elm83}.
\citet{elm94} argued that this spacing of giant \ion{H}{2} regions
is attributable to gravitational collapse of
{\it magnetized} gas within spiral arms.

An origin of spur formation with gravitational instability in 
spiral arms is natural because it relies only upon the fundamental 
physical agents known to be present:
galactic differential rotation, compression by the spiral pattern,
and self-gravity; we shall show that well-developed
spurs may require magnetic fields as well. 
Two barriers to gas condensations on a large scale are 
insufficient surface density and excessive shear;
because both of these impediments are reduced simultaneously
in spiral arms (e.g., \citealt{rob69,elm87a}), these regions are
prone to self-gravitating instabilities that can
produce nonaxisymmetric structure.
Condensations may also grow inside
spiral arms under the combined action of the Parker and
gravitational instabilities, subsequently being
carried downstream by galactic differential rotation 
to appear as spurs.
In this paper, however, we isolate and concentrate only on 
local gravitational instability occurring in two-dimensional (2D),
razor-thin disks; the effects of the Parker instability 
will be studied in future work.

Several previous studies have considered self-gravitating growth of 
perturbations within spiral arms in the {\it linear} regime.
\citet{bal85} analyzed local stability of quasi-axisymmetric 
disturbances (i.e.\ $\mathbf{k}$ perpendicular to the local
spiral arm), and showed that in the linear regime growth is
intrinsically transient, stabilized by the expanding,
shearing flow as the medium leaves the arm region.
\citet[hereafter B88]{bal88} analyzed local amplification of
shearing wavelets with arbitrary initial local wavenumber,
allowing for a varying background shear profile self-consistent
with the varying ISM compression in the arm/interarm regions.
\citet{elm94} incorporated effects of magnetic fields in a WKB
analysis of instantaneous growth in a spiral arm region,
without explicitly taking into account the expansion
and varying shear flow downstream (relevant to the 
swing mechanism; see B88).

Of previous work, the most detailed study relevant to the 
problem of spiral arm spur formation is that of B88. He showed that
transient gravitational instabilities within compressed regions
have two preferred directions for initial wavenumbers:
nearly parallel and nearly perpendicular to the arm.
He also suggested that closely-spaced periodic spurs may naturally
develop from the former, while the latter could produce 
(non-periodic) spurs under certain conditions for the background
velocity field.

In this paper, we investigate {\it nonlinear} evolution of local 
self-gravitating perturbations in the magnetized, gaseous ISM as they 
interact with the spiral potential in a differentially-rotating
galaxy. Our study extends the work of B88 by including the
effects of magnetic fields and nonlinearity.
It extends our own previous nonlinear study of self-gravitating
evolution in shearing, magnetized, featureless disks \citep{kim01}
to incorporate the strong density and velocity gradients in the 
equilibrium gas profile driven by a stellar spiral arm potential.
For thin, featureless disks, there are two different mechanisms that
grow density structure:
the magneto-Jeans instability when shear is weak or magnetic fields 
are strong; and the magnetically-modified swing amplifier when
shear is strong and the magnetic field is moderate or weak.
Here, we study the modifications of these mechanisms by the
presence of the spiral arm.

We proceed technically by considering a local 
patch of an infinitesimally thin
gaseous disk subject to an external (stellar plus dark matter) potential.
The model disk is assumed to be isothermal and
initially magnetized parallel to the spiral arms.
We adopt \citet{rob69}'s coordinate system having two orthogonal
axes parallel and perpendicular to the gradient in the local external spiral 
potential. We first obtain one-dimensional (1D) density and velocity
profiles with variation only in the direction perpendicular to the
arm. We then apply small perturbations and follow their 
development with 2D direct numerical 
simulation. For comparison with these simulations, we also
integrate the magnetized equivalent of the linear shearing-wavelet
equations of B88.

The remainder of this paper is organized as follows:
In \S 2, we introduce the Roberts coordinate system, present the vertically
integrated, 2D magnetohydrodynamic (MHD) 
equations in these spiral-arm coordinates, and describe
our simulation model parameters and numerical method.  
In \S 3, we pursue 1D spiral shock equilibria under the given background 
conditions, and study quasi-axisymmetric instabilities.
We present in \S 4 a set of MHD simulations with zero, sub-equipartition, 
and equipartition magnetic fields, to explore 2D structure formation.
We analyze density and velocity structures of the spurs that
we find form, and compare these structures with analytic predictions. 
We present density snapshots from late stages of evolution for
several models, showing that spurs frequently fragment to produce
dense knots with masses comparable to giant clouds or massive
young stellar associations.
We summarize our results and discuss the conclusion of present work in \S5. 
In Appendix A, we present the linear-theory perturbation equations for 
magnetized flows in spiral arms.

\section{Basic Equations and Model Parameters}

\subsection{Coordinates and Basic Equations}

We study the nonlinear response of a self-gravitating, magnetized,
differentially rotating, razor-thin,
gaseous disk to an externally-imposed spiral potential.
The spiral potential, which is thought of as arising
primarily from the stellar component, is assumed to be 
tightly wound with a pitch angle $i=\arctan{(\lambda_R/\lambda_\theta)} \ll 1$
and rigidly rotating relative to an inertial frame 
at a constant pattern speed $\Omega_p$. 
The spiral forcing tends to induce noncircular
motion in gaseous disks that would otherwise
rotate with a angular velocity $\Omega(R)$ in the azimuthal
direction $\hat{\theta}$, where $R$ is the
galactocentric radius. The induced radial velocities are generally
only a few percent of $R\Omega(R)$, but enough to
compress gas into a highly nonlinear state.
We investigate the gravitational instabilities that may develop
in these highly-compressed spiral-arm regions.

Following \citet{rob69} and B88, we construct a 
local Cartesian frame whose center
lies at a galactocentric distance $R_0$ and 
rotates at $\Omega_p$ about the galactic center.
The two coordinates correspond to radial $(R-R_0)$ and
angular ($R_0(\theta-\Omega_pt)$) displacements with respect to
the center of our moving frame.
We then tilt our frame by an angle $i$ 
such that $\hat{x}$ and $\hat{y}$ correspond to 
the directions perpendicular and parallel to the local
spiral arm, respectively, and consider a local
rectangular box with size $L_x\times L_y$, 
as depicted in Figure \ref{arm_coord}.
We expand the compressible, ideal MHD equations in the rotating
frame, and take the approximations that 
$\sin i \ll 1$ (tightly wound arms), 
$x, y \ll R_0$ (local model),  and 
the velocities induced by the spiral arm are much smaller
than $R_0\Omega(R_0)$.
Neglecting the terms arising from the curvilinear
geometry and integrating the resulting equations in the 
vertical direction, we obtain the following set
of 2D equations:
\begin{equation}\label{con}
  \frac{\partial\Sigma}{\partial t} + \nabla\cdot(\Sigma \mathbf{v}_T) = 0,
\end{equation}
\vspace{-0.8cm}
\begin{mathletters}
\begin{eqnarray}
  \frac{\partial\mathbf{v}}{\partial t} +
        \mathbf{v}_T\cdot\nabla\mathbf{v}
    & = & -\frac{1}{\Sigma}\nabla\Pi
       + \frac{1}{4\pi\Sigma}(\nabla\times\mathbf{B})\times\mathbf{B} 
         \nonumber \\
      & & + q_0\Omega_0 v_x \mathbf{\hat{y}}
       - 2\mathbf{\Omega}_0\times\mathbf{v}
       - \nabla (\Phi_g + \Phi_{\rm ext}), 
\eqnum{2}
\end{eqnarray}
\end{mathletters}
\vspace{-0.5cm}
\begin{equation}\label{ind}
  \frac{\partial \mathbf{B}}{\partial t} =
       \nabla\times(\mathbf{v}_T\times\mathbf{B}),
\end{equation}
\begin{equation}\label{Pos}
   \nabla^2\Phi_g = 4\pi G \delta(z)\Sigma,
\end{equation}
and
\begin{equation}\label{eos}
\Pi = c_s^2\, \Sigma,
\end{equation}
(cf, \cite{rob69,rob70,shu73,bal85}; B88). 
Here, $\Sigma$ is the surface density, 
$\Pi$ is the 2D vertically-integrated pressure,
$\Phi_g$ is the self-gravitational potential, and
$\mathbf{B}$ is the midplane value of the 3D magnetic field
times the square root of the unperturbed ratio of surface density 
to midplane volume density. 
The total velocity as viewed in the rotating frame is
denoted by $\mathbf{v}_T$, while $\mathbf{v}$
represents the part induced by the external
spiral potential $\Phi_{\rm ext}$;
that is, $\mathbf{v} \equiv \mathbf{v}_T - \mathbf{v}_0$, where
$\mathbf{v}_0$ is the velocity arising from galactic 
rotation\footnote{By direct coordinate transformation, 
$\mathbf{v}_0 = 
[R_0(\Omega_0-\Omega_p) - q_0\Omega_0(x\cos i-y \sin i)]
(\sin i\,\mathbf{\hat{x}}+ \cos i\,\mathbf{\hat{y}})$.
Within the local, tightly-winding limit ($x, y \ll R_0$ 
and $\sin i \ll 1$) that we adopt,
the background circular velocity reduces to 
$\mathbf{v}_0 = R_0(\Omega_0-\Omega_p) \sin i\,\mathbf{\hat{x}}
+ [R_0(\Omega_0-\Omega_p) - q_0\Omega_0 x]\mathbf{\hat{y}}$.
This form of the background velocity permits local shearing periodic
boundary conditions.}.
In equation (2), 
$\Omega_0 \equiv \Omega(R_0)$ and $q_0 \equiv -(d\ln\Omega
/d\ln R)|_{R_0}$ measures shear rate in the background flow 
in the absence of a spiral pattern,
while $\cs$ in equation (\ref{eos}) is the isothermal
sound speed. In terms of $q_0$ and $\Omega_0$, the local
epicyclic frequency $\kappa_0$ is given by 
$\kappa_0^2 \equiv R^{-3}d(R^4\Omega^2)/dR|_{R_0} = (4-2q_0)\Omega_0^2$.
Finally, $G$ and $\delta$ in equation (\ref{Pos}) are the
gravitational constant and the Kronecker delta, respectively.

As expressed by equation (\ref{eos}), in this paper we adopt an
isothermal equation of state.
In equations (2) and (\ref{ind}), we treat 
the effective scale height of the magnetic field distribution 
as a constant in both space and time. 
In the present height-integrated formulation, 
our dynamical model does not 
capture the potential consequences of Parker instability and
magnetorotational instabilities for disk evolution (cf. \citealt{kim01}).
Nevertheless, because our equations contain the essential physical 
ingredients involved in self-gravitational instabilities 
inside spiral arms, the present work represents the first step
in a more comprehensive three-dimensional study.

To complete the governing equations, 
we must specify the external spiral
potential $\Phi_{\rm ext}$. 
For compatibility with our local model, 
$\Phi_{\rm ext}$ must be periodic in $x$ and independent of $y$;
we adopt the following simple form:
\begin{equation}\label{ext_P}
\Phi_{\rm ext} = \Phi_0 \cos \left(\frac{2\pi x}{L_x}\right),
\end{equation}
which is a local analog of a logarithmic potential of 
\citet{rob69} and \citet{shu73}. 
Since $x$ varies from $-L_x/2$ to $L_x/2$,
$\Phi_{\rm ext}$ attains its minimum at the center ($x=0$)
for $\Phi_0 <0$. The distance $L_x$ is the arm-to-arm separation,
equal to $2\pi R_0\sin i/m$ for an $m$-armed spiral.

\subsection{Model Parameters}

In this paper, we adopt an isothermal equation of state; for scaling
our solutions, we shall use an effective isothermal speed of sound
$\cs=7{\,\rm\,km\,s^{-1}}$, corresponding to a
mean thermal pressure $P/k 
\sim 2000-4000{\,\rm K\,cm^{-3}}$ \citep{hei01}
and mean midplane density $n_{\rm H} \sim0.6\,{\rm cm^{-3}}$ \citep{dic90}. 
For the spiral potential parameterization, we take pattern speed
$\Omega_p = \Omega_0/2$ and pitch angle $\sin i =0.1$, respectively.
For a two-armed spiral pattern located at distance $R_0=10$ kpc from
the galactic center, we thus have $L_x = \pi R_0 \sin{i} = 3.1$ kpc.
In the solar neighborhood, 
the angular velocity of the Galactic rotation is
$\Omega_0= 26 {\,\rm km\,s^{-1}\,kpc^{-1}}$ \citep{bin87},
and $q_0\thickapprox1$ for a near-flat rotation,
so that $\kappa_0\thickapprox2^{1/2}\Omega_0$.
The corresponding orbital period is
$\torb \equiv  2\pi/\Omega_0 =  2.4\times 10^8{\,\rm yr\,} (\Omega_0/
26\,{\rm km\,s^{-1}\,kpc^{-1}})^{-1}$. 
The size of the simulation domain in
the $y$-direction is chosen to be $L_y=2L_x$
(we find that our results are generally independent of this choice).

In the absence of external potential perturbation,
our model disks have uniform surface density $\Sigma_0$, 
and uniform magnetic field $\mathbf{B}_0$ that 
points in the $\hat{y}$-direction. 
To characterize $\Sigma_0$ and $B_0$, we introduce two
dimensionless parameters,
\begin{equation}\label{Q0}
Q_0 \equiv \frac{\kappa_0 c_s}{\pi G \Sigma_0}
\end{equation}
{\rm and}
\begin{equation}\label{beta}
\beta_0\equiv \frac{c_s^2}{\vA^2}
= \frac{4\pi\rho_0 c_s^2}{B_{\rm 0,3D}^2},
\end{equation}
where  the \Alf speed is defined by
$\vA \equiv B_0^2/4\pi\Sigma_0\equiv B_{\rm 0,3D}^2/4\pi\rho_0$ 
with 3D field strength $B_{\rm 0,3D}$ at the disk midplane.
The Toomre $Q_0$ parameter is a measure of the 
axisymmetric stability
of a rotating, unmagnetized disk of {\it uniform} density.
For the background conditions adopted in this paper, the 
corresponding mean gas surface density and mean
magnetic field strength are
\begin{displaymath}
\Sigma_0 = \frac{19}{Q_0} \,\Msun\pc^{-2}
 \left(\frac{\cs}{7.0 {\rm\,km\,s^{-1}}} \right)
        \left(\frac{\kappa_0}{36\,{\rm km\,s^{-1}\,kpc^{-1}}}\right),
\end{displaymath}
and 
\begin{displaymath}
B_{\rm 0,3D} = \frac{3.2}{\sqrt{\beta_0}} \, \mu\,{\rm G}
\left(\frac{n_{\rm H}}{1 {\rm\,cm^{-3}}}\right)^{1/2}
\left(\frac{\cs}{7.0 {\rm\,km\,s^{-1}}} \right).
\end{displaymath}
Note that the total mass contained in the box is then
roughly
$M_{\rm tot} = L_xL_y\Sigma_0 \sim (4/Q_0)\times 10^8 \,\Msun.$

Finally, the strength of the external spiral potential is
characterized by 
\begin{equation}\label{F_arm}
F\equiv \frac{2}{\sin i}\left(\frac{|\Phi_0|}{\Omega_0^2 R_0^2}\right),
\end{equation}
which measures the amplitude of the perturbed radial force 
$2\pi|\Phi_0|/L_x$ as a fraction
of the mean axisymmetric gravitational force field \citep{rob69}.
We vary $Q_0$, $\beta_0$, and $F$ to represent disks in various 
physical conditions, while fixing the other input parameters 
as described above.
Note that other choices of $\Omega_0$, $\cs$, $L_x$, and $\sin i$
may be substituted for our fiducial values, provided that
the ratio $\cs\Omega_0/L_x$ remains unchanged, and $F$ is
interpreted as varying proportional to $\sin i$.

\subsection{Numerical Method}

The nonlinear evolution of gaseous model disks is followed by
integrating the governing set of equations (\ref{con})$-$(\ref{eos})
using a modified version of the ZEUS code 
originally developed by \citet{sto92a, sto92b}. 
ZEUS is a time-explicit, operator-split, finite-difference method
for solving the MHD equations 
on a staggered mesh. ZEUS employs ``constrained transport''
to guarantee that $\nabla\cdot\mathbf{B}=0$ within machine
precision, and the ``method of characteristics'' for accurate
propagation of Alfv\'enic disturbances \citep{eva88, haw_sto95}.
For this work we implement
shearing box boundary conditions, 
in which the $y$-boundaries are perfectly periodic and 
the $x$-boundaries are shearing-periodic \citep{haw95}.
For gaseous self-gravity, we implement a 
shearing-sheet Poisson solver \citep{gam01} via FFTs such 
that $\Phi_k = -2\pi G\Sigma_k/|k|.$ 
For less diffusive advection of hydrodynamic variables, 
we apply the velocity decomposition method which treats
the contribution from the background shearing parts
as source terms. In order to minimize the errors arising from
discontinuities in the flow characteristics across
the $x$-boundaries, the ghost zones adjoining the $x$-boundaries
are kept active. 
Our implementation of the ZEUS code has been
fully tested on a variety of test problems and used for the 
study of gravitational instabilities in galactic disks without
spiral-arm features \citep{kim01}. For the interested reader, we refer
to \citet{kim01} for a detailed description of the code and 
its test results. Our resolution for the 2D simulations presented 
in this paper is 256$\times$512 zones. 

\section{One-Dimensional Spiral Shocks and Their Stability}

It is straightforward to use our numerical code to obtain
1D solutions for the gaseous response to an
applied spiral potential. In this section, we present
radially-localized solutions for
steady-state 1D spiral shock structures
and investigate their stability  by performing fully nonlinear,
time-dependent calculations. 
Steady-state 1D solutions including shocks were originally obtained
by \citet{rob69}, \citet{rob70}, \citet{shu72}, \citet{shu73},
and \citet{tom87} 
for non-self-gravitating flows,
and by \citet{lub86} for viscous, self-gravitating, unmagnetized flows.
\citet{woo75} followed a time-dependent approach and 
showed that spiral shocks indeed form within one or two crossing times
of gas through the spiral density waves. 
While his treatment allowed for the effects of ultraharmonic 
resonances (cf.\ \citealt{shu73}) in driving the nonlinear gaseous response,
\citet{woo75} did not address the stability of spiral shocks 
when self-gravity and magnetic fields are incorporated.
Similar works for 2D galactic shocks including both
spiral and bar potentials 
were reported by \citet{rob79} and \citet{van81}.

Linear analyses of 1D gravitational instabilities 
of the ISM inside spiral arms were 
performed by \citet{bal85};
here we shall present 
the results of nonlinear simulations.
\citet{bal85} focused on the development histories of
gravitationally-amplified perturbations, showing that in linear
theory the background expansion of the flow as it exits the arm 
limits growth.
Thus, in analogy to the situation for nonaxisymmetric 
perturbations in a shearing background \citep{gol65},
there are no true linear instabilities but only a transient
growing phase, analogous to the swing amplifier. 
Just as nonlinear effects nevertheless lead to an instability
threshold for the swing amplifier \citep{kim01}, one may
expect a nonlinear instability threshold in the 1D spiral
shock problem. In this section,
we concentrate on finding critical $Q_0$ values for the
existence of equilibrium spiral shocks for given
galactic conditions. 
As we vary only the magnetic field strength and the spiral perturbation
amplitudes, our coverage of the parameter space is of course incomplete. 
The trends shown in the critical $Q_0$ curves are interesting
in themselves, but are also important for delimiting the portion
of parameter space for which nonaxisymmetric studies are warranted.
One-dimensional equilibrium shock profiles constructed in this
section will serve as initial conditions for our 
nonaxisymmetric simulations in \S 4.

We begin by considering a non-self-gravitating medium with
uniform density $\Sigma_0$, background shear profile $\mathbf{v}_0$,
and uniform magnetic field characterized by $\beta_0$.
We then impose  an external spiral perturbation and 
slowly increase its amplitude up to a desired level, $F$ (eq.\ [\ref{F_arm}]).
The spiral forcing drives magnetosonic waves, 
but these waves with $\cs/v_{0,x} < 1$ soon steepen,
eventually developing a shock front.
The surface density first overshoots and oscillates about the 
corresponding steady-state solution for given $\beta_0$ 
and $F$, and then gradually 
converges to it. The typical time to reach a stationary
state is about 5 orbital times. 
We check the code accuracy by comparing the numerical solutions with 
published solutions
obtained from solving, with a shock-fitting procedure, the  
time-independent ordinary differential equations 
\citep{rob69,shu72,shu73}. 

We then slowly turn on the self-gravitational force, as we did
for $F$, to obtain a self-gravitating equilibrium profile corresponding
to a given $Q_0$.
Two examples of stable equilibria for parameter values 
$Q_0=2.0$, $\beta_0=10$, $F=3\%$
and $Q_0=1.5$, $\beta_0=1,$ $F=3\%$
are respectively illustrated in Figures \ref{1Dshock10}
and \ref{1Dshock1} (solid curves).
Note that gas is flowing from left to right.
Dotted curves in Figures \ref{1Dshock10}$a$-\ref{1Dshock10}$e$ 
and \ref{1Dshock1}$a$-\ref{1Dshock1}$e$ indicate
the unperturbed state with $F=0$, while the dashed lines in 
Figures \ref{1Dshock10}$b$ and \ref{1Dshock1}$b$ mark the sound speed.
For comparison, we also plot the surface density profiles for
the non-self-gravitating counterparts as dashed lines 
in Figures \ref{1Dshock10}$a$ and \ref{1Dshock1}$a$ .
A well-defined shock front occurs at $x/L_x \approx -0.02\; (-0.05)$ with
a thickness $\Delta x /L_x \approx 0.03\; (0.02)$  for the 
$\beta_0=10 \;(1)$ case.
Across the shock front, $v_x$ experiences a sharp 
deceleration from supersonic to subsonic magnitude, while
the gas and magnetic field are both compressed. 
After the density peak, gas begins to expand and $v_x$ becomes again 
supersonic after crossing another sonic point at 
$x/L_x \approx 0.079\;(0.086)$ for $\beta_0=10 \;(1)$.
For flows with one independent spatial variable, 
mass and magnetic flux conservation guarantees 
$\Sigma \propto B \propto v_x^{-1}$, 
so that
$\beta \propto v_x$, as confirmed by Figures 
\ref{1Dshock10}$b$,\ref{1Dshock10}$d$ and 
\ref{1Dshock1}$b$,\ref{1Dshock1}$d$. 
The conservation of the potential vorticity\footnote{
In these expressions for $\xi$, the curl operator and velocities
are defined in the local Cartesian frame built in \S 2.1.
When expressed in standard cylindrical coordinates retaining
curvilinear terms, $\xi=|\nabla\times \mathbf{V}|/\Sigma =
|\nabla\times \mathbf{v}_T + 
2\mathbf{\Omega}_p|/\Sigma$, where $\mathbf{V}$ is the fluid
velocity seen in the inertial frame and $\mathbf{v}_T\equiv
\mathbf{V} - R\Omega_p\mathbf{\hat{\phi}}$.}
$\xi\equiv |\nabla\times\mathbf{v}_T+2\mathbf{\Omega}_0|/\Sigma =
|\nabla\times\mathbf{v}+(2-q_0)\mathbf{\Omega}_0|/\Sigma$
\citep{hun64,gam96,gam01} leads to an equilibrium profile 
$Q=Q_0(\Sigma/\Sigma_0)^{-0.5}$ for isothermal gas 
or $Q=Q_0\gamma^{1/2}(\Sigma/\Sigma_0)^{\gamma/2-1}$ for
polytropic law $\Pi \propto \Sigma^\gamma$ \citep{bal85},
and implies that the local shear rate varies as 
\begin{equation}\label{loc_shear}
q\equiv -\frac{\partial \ln{\Omega}}{\partial\ln R}\bigg|_{R_0}
= -\frac{1}{\Omega_0}\frac{dv_{y,T}}{dx }
=2-(2-q_0)\frac{\Sigma}{\Sigma_0},
\end{equation}
 \citep{kim01}.
As a consequence, shear is reversed where $\Sigma/\Sigma_0 > 2$
for $q_0=1$, as shown in Figures \ref{1Dshock10}$c$
and \ref{1Dshock1}$c$.

Our simulation domain contains an
ultraharmonic resonance where $v_{x,0}^2-\cs^2=
(L_x/2\pi)^2\kappa^2/n^2$ 
for a non-self-gravitating, unmagnetized medium,
with integer $n$ denoting the order of the resonance \citep{shu73}.
Using the parameters adopted in \S2.2,
one can show that
$x/L_x=0.245$ corresponds to a second harmonic resonance ($n=2$).
Although this resonance is easily detectable when spiral
arm forcing is relatively weak \citep{shu73}, 
the local maximum around $x/L_x\approx0.25$ in 
the dashed curve for the non-self-gravitating model
in Figure \ref{1Dshock10}$a$ still 
traces the (weak) second harmonic resonance.
Embedded (strong) magnetic fields and self-gravity significantly
change the resonant conditions (cf.\ \citealt{lub86}),
so that $\beta_0=1$ models (e.g., Fig.\ \ref{1Dshock1}$a$) and
strongly self-gravitating $\beta_0=10$ models (e.g., solid line
in Fig.\ \ref{1Dshock10}$a$) do not show any indication of
the resonance.

In general, self-gravity tends to move the shock 
front downstream, enhance the maximum surface
density, and widen the shock front\footnote{Since the 
gravitational potential is continuous across the shock front,
self-gravity does not affect the shock jump conditions (cf.\
\citealt{shu92}). The peak density is not always reached 
immediately behind the shock \citep{shu73}.
However, the equilibrium profiles presented in Figures \ref{1Dshock10} 
and \ref{1Dshock1} could not resolve the separation between the 
shock front and the locus of the
maximum density.}.
\citet{lub86} found that with viscosity included (so as to
represent a nonzero mean free path in a cloud-fluid), the gas
self-gravity suppresses the tendency of the gas to form a shock
when the gas content is large.
We find the similar result that when the spiral arm
forcing is relatively weak (i.e., the gaseous self-gravity
exceeds the force due to stars), the shock disappears, leaving quite
symmetric density configurations
(see, for example, model H1 in Fig.\ \ref{HDevol}$a$).
Notice the magnitudes of $\Phi_{\rm ext}$ and $\Phi_g$ are
comparable in Figures \ref{1Dshock10}$f$ and  \ref{1Dshock1}$f$,
although $\Phi_g$ has larger gradients.

With smaller $Q_0$ and/or larger $F$ (i.e., larger self and/or external
gravity), spiral shock fronts move further downstream, and 
the arm-to-interarm contrast in the stationary density profile grows,
increasing the susceptibility to gravitational instability.
For models with $Q_0$ less than some critical value
(for other parameters fixed),
no stationary configuration can be found. This is either because
the equilibrium becomes nonlinearly unstable for sufficiently strong 
self-gravity, or because
no equilibrium for a given background condition
is compatible with the chosen pattern speed
and strength of the spiral potential.
These critical $Q_0$ values 
together with corresponding local values of $Q$ at the density peak,
$\Qsp\equiv Q_0(\Sigma_{\rm max}/\Sigma_0)^{-1/2}$,
are indicated in Figure \ref{Qcrit}.

Gravitationally unstable cases, as represented by filled symbols in 
Figure \ref{Qcrit}, occur when the magnetic field and external forcing 
are weak. Stronger magnetic fields provide pressure support against
gravitational collapse if $Q_0$ is not as small.
As $Q_0$ is lowered, the equilibrium shock front moves gradually 
downstream, where the gradient of spiral potential is smaller.
Beyond a certain point (when the shock approaches the maximum of 
$\Phi_{\rm ext}$), the shock front is no longer stationary but 
moves back and forth, with rapidly changing density.
This implies that steady-state solutions of equations 
(\ref{con})$-$(\ref{eos})
do not exist for a given background state.
The boundaries of cases with this sort of behavior are marked by 
open symbols in Figure \ref{Qcrit}.
Note that the critical $Q_0$ values, $\Qc$, for purely hydrodynamic models 
all represent instances of gravitational instabilities, ranging over
$\Qc\sim 1.0-2.8$ for $F \leq6\%$. As $\beta_0$ decreases (stronger
mean magnetic fields), the range of $\Qc$ also decreases, 
giving $\Qc\sim 0.9-2.0$ for $\beta_0=10$
and $\Qc\sim0.7-0.9$ for $\beta_0=1$.

In cases that are gravitationally unstable, 
perturbations grow as they move downstream off the arm.
For this gravitational runaway to occur,
amplification of any 
perturbations must be high enough to drive the system
into the nonlinear regime before the onset of stabilization 
associated with physical expansion and high interarm $Q$.
\cite{bal85} suggested that $\Qsp$ is {\it the} parameter that 
characterizes the gravitational response of gas to applied
perturbations. For marginally stable models with 
$1\%\leq F \leq 6\%$,
Figure \ref{Qcrit} shows that 
$\Qsp\sim0.73-0.98$ for $\beta_0=\infty$, 
which is in good agreement with 
\citet{bal85}'s finding that $\Qsp < 1$ is required (but not 
sufficient) for significant growth of perturbation. 
The main reason why $\Qsp<1$ does not necessarily produce axisymmetric
instability is of course that the background surface
density is not uniform; at least a region of the shortest unstable
wavelength must have $Q<1$.
For cases where marginality represents incompatibility of equilibria
with prescribed background conditions,
$\Qsp$ converges to 0.5 for $\beta_0=10$ and to 0.4
for $\beta_0=1$ as $F$ increases. Note that $\Qsp$ does not
include the stabilizing effect of magnetic fields;
including magnetic fields increases the effective $Q$ by a factor
$(1+\Sigma/(\beta_0\Sigma_0))^{1/2}$, which is generally larger 
than unity at the limiting value of $\Qsp$.

\section{Two-Dimensional Simulations}

In the previous section, we discussed the nonlinear stability of the gas 
flow through spiral arms to perturbations in which all spatial gradients
point in the direction transverse to the spiral arm 
($\hat{x}$-direction). 
These may be thought of as quasi-axisymmetric instabilities. We now
generalize to study the development of perturbations in which
gradients may point in any direction with respect to the 
spiral arm. Disturbances where the gradients lie primarily along
the arm ($\hat{y}$-direction) may be thought of as
quasi-azimuthal (nonaxisymmetric) perturbations.

For the initial conditions of our 2D simulations, we use the 
equilibrium shock profiles 
obtained from 1D calculations in \S 3.
We consider cases with arm-interarm contrast in surface density
less than a factor 12.
Initial perturbations are realized by a Gaussian random density
field with flat power for $1 \leq kL_x/2\pi \leq 64$ and
zero power for $64 < kL_x/2\pi$.
For the amplitude of perturbations,
we measure the standard deviation $\epsilon_0$ of the initial
density fluctuation in real space, and fix $\epsilon_0=1\%$.
These adopted density perturbations do not attempt to represent 
realistic interstellar perturbations:
in real galaxies, perturbations may have power-law spectra in both
velocity and density fields with significantly larger amplitudes. 
The chosen amplitude and spectrum, nevertheless, allow us 
to monitor evolutionary behavior in both linear and nonlinear regimes,
and seed the most dominant nonaxisymmetric mode of 
the instability in the simplest possible way.

The model parameters and nonaxisymmetric simulation results we present in
this section
are summarized in Table \ref{tbl-1}. Column (1)
labels each run. Columns (2) and (3) list the two basic 
parameters of our model disks: the Toomre
stability parameter $Q_0$ and the plasma parameter $\beta_0$ 
characterizing the magnetic field strength (see eqs.\
[\ref{Q0}] and [\ref{beta}]). The amplitude of 
spiral arm forcing is given in terms of $F$  
(see eq.\ [\ref{F_arm}]) in column (4).
The peak surface density $\Sigma_{\rm max}$ and 
local $\Qsp$ of the 1D, stationary solution are
listed, respectively, in columns (5) and (6), while 
column (7) gives 
the arm width determined at $\Sigma=(\Sigma_{\rm max} +
\Sigma_{\rm min})/2$. 
In the cases where spurs form,
their mean spacing ($\lambda_y$) measured along the arm is given
in columns (8) and (9) in the units of 
the arm-to-arm distance $L_x$ and the arm width $W$, respectively. 
Finally, column (10) gives the corresponding normalized wavenumber 
$\Kymax \equiv \ljsp/\lambda_y
= 2\pi (\lambda_y\kjsp)^{-1}$ of spurs, 
where the local Jeans wavenumber $\kjsp$
at the density peak is defined by
\begin{equation}\label{k_Jsp}
\kjsp\equiv \frac{2\pi}{\ljsp} =\frac{2\pi G \Sigma_{\rm max}}{\cs^2}.
\end{equation}
Note, for comparison, that 
magneto-Jeans instability in a {\it uniform} disk with 
embedded magnetic fields would give $\Kymax = \case{1}{2}$ for
$\beta_0 \ll 1 $ and $\Kymax \sim \case{1}{2}-\case{3}{4}$ 
for $\beta_0 \simgt 0.1 $, while swing amplification 
generally favors $\Kymax \sim 0.15-0.4$ for $\beta_0 \simgt 1$
\citep{kim01}. 

\subsection{Purely Hydrodynamic Models}

For purely hydrodynamical simulations, 
we select models that are very close to
marginal stability for a given arm strength $F$. Note that the
critical $Q_0$ values for $\beta=\infty$ are 1.67, 2.01, and 2.26,
for $F=1$\%, 2\%, and 3\%, respectively.
In Figure \ref{HDevol}, we display the initial surface density profiles
and the evolutionary histories of maximum surface densities for
model H1$-$H3.

Generally speaking, {\it unmagnetized} configurations that are
quasi-axisymmetrically
stable are found to be stable also to nonaxisymmetric perturbations,
unless $Q_0$ and $F$ are quite small.
In models H2 and H3, with a realistic arm-interarm contrast in
surface density of
$\Smax/\Sigma_{\rm min}=6.2$ and 11.5, respectively, 
the nonaxisymmetric growth of density is so mild that no gravitationally
bound structure forms within the simulation interval.
The two main mechanisms that can produce 
local growth of perturbations in unmagnetized, 
razor-thin disks are quasi-axisymmetric instability and swing amplification
(see B88, who refers to the quasi-axisymmetric modes as Jeans instability). 
Since our chosen initial conditions for nonaxisymmetric simulations are 
already stable to 1D quasi-axisymmetric instability, one can only expect 
perturbations to grow as wave crests swing from leading to trailing
configurations. Figures \ref{HDevol}$b$ 
shows that swing amplification in models H2 and H3 is in fact very weak,
producing only small-amplitude fluctuations in the density.
This is because the high density compression in spiral arm regions
produces reversed shear (see eq.\ [\ref{loc_shear}] with $q_0=1$
and Figs.\ \ref{1Dshock10}$c$, \ref{1Dshock1}$c$) 
under which condition classical swing amplification is
essentially shut off. Another viable mechanism within spiral arms
is {\it divergence} swing amplification that allows 
growth of leading wavelets in expanding background flows (B88).
It is, however, very difficult to supply leading perturbations
to spiral arms in a natural way, 
because the perturbations carried in from interarm
regions are preferentially trailing.
In interarm regions where $\Sigma/\Sigma_0<2$ so background shear is in the
same sense as epicyclic motion, on the other hand, 
$Q$ is high enough to suppress the classical swing mechanism. 
For example, model H3 has $Q \approx 2.8$ at $x/L_x = 0.3$. 

Because of its smaller $Q_0$ and flatter density distribution,
model H1 experiences stronger swing amplification than models H2 and H3.
With smaller $F$, self-gravity smears out the spiral shock
and makes the equilibrium profile more symmetric. 
Lower $\Smax$ implies the sense of shear is ``normal'' in the major
portion of a disk.
Still, as can be seen in Figure \ref{HDevol}$b$, 
growth of perturbations occurs so slowly that it requires four
successive passages through spiral arm regions to amplify them
significantly. Had we begun the evolution with higher
perturbation amplitudes, less time would be required to attain 
a fully nonlinear state.
As a result of swing amplification, model H1 forms three (weak) spurs
in the direction perpendicular to the spiral arm,
with an average separation of $\sim 2.1$ kpc.
When allowed to evolve further, they fragment and collapse.

The spurs that form in model H1 are shown in Figure \ref{HDimage},
where we plot the density at $t/\torb=4.7$, and for comparison 
the density in the stable model H3 at $t/\torb=4.0$.
Warped by the nonlinear background shear and expansion flow off the arm,
the local wavefronts defining the spurs in model H1 are well described 
by the kinematic formula provided by B88. Namely, the local tangent to
the wavefront is given by $dy/dx = -\kx/\ky = \mathcal{T}$ with
\begin{equation}\label{Wfront}
\mathcal{T} \equiv \frac{1}{\mathcal{R}}\left[
  \frac{\kappa_0^2\Smax}{2\Omega_0^2\Sigma_0}\tau
  - 2\int_0^\tau \mathcal{R} d\tau - \frac{\kx(0)}{\ky}\right],
\end{equation}
where $\mathcal{R}\equiv \Smax/\Sigma=v_{x,T}/v_{x,T,\rm min}$ is
the local surface density expansion factor,  
$\tau \equiv \Omega_0
\int_{x_{\rm sp}}^x v_{x,T}^{-1} dx$ is a dimensionless elapsed
time that is measured from the shock location (or density peak), $x_{\rm sp}$,
$\ky=2\pi/\lambda_y$, where $\lambda_y$ is the spur spacing,
and $\kx(0)$ is the local $x$-wavenumber at 
$\tau = 0$. The function $\mathcal{T}$ is defined in terms
of the unperturbed state. From mass conservation, we can write 
the time element in terms of the expansion factor $\mathcal{R}$ as 
\begin{equation}\label{tele}
d\tau = \frac{\Omega_0\Smax dx}{R_0(\Omega_0-\Omega_p)\sin i\Sigma_0
\mathcal{R}},
\end{equation}
so that $\mathcal{T}$ only depends on the variation of surface density with 
$x$ via $\mathcal{R}(x)$. Using equation (\ref{tele}), one can show that
the second term in the square brackets in  equation (\ref{Wfront}) 
is directly proportional to $x-x_{\rm sp}$; the first term, 
from the definition of $\kappa_0^2$, equals $(2-q_0)\tau\Smax/\Sigma_0$.
The curves whose local gradient is defined by
$\cal{T}$ with $\kx(0)/\ky=-2.2$ are overlaid on the spurs 
in Figure \ref{HDimage}, showing excellent agreement with
the results of numerical simulations. 

Our numerical evidence of the
relative stability of purely hydrodynamic systems to nonaxisymmetric
perturbations might seem in conflict 
with B88's conclusion from linear theory that nonaxisymmetric wavelets
can grow rapidly within $\sim0.3$ orbital times. 
In fact, our results are entirely consistent with the predictions
of B88's general framework, but our numerical models explore a 
different part of the parameter space of background conditions --
especially $\Qsp$ -- from that explored by B88. 
In particular, B88 considered a very soft 
polytropic equation of state ($\gamma=0.5$), such that he could easily 
achieve $\Qsp=Q_0\gamma(\Smax/\Sigma_0)^{\gamma/2-1}\sim0.6$ from solar 
neighborhood conditions, while our
isothermal ($\gamma=1$) models have $\Qsp>0.9$ as Table \ref{tbl-1} indicates.
As Figure 5 in B88 shows, amplification factors are very
sensitive to $\Qsp$, dropping by about one order of magnitude
as $\Qsp$ increases from 0.52 to 0.67. Naive extrapolation 
of this result suggests that models with $\Qsp>0.9$ may remain stable; 
we have indeed verified that the solutions of the corresponding linear
equations (see Appendix A for magnetized versions) show very low
amplifications for our background models.
Because our 2D simulations require an initial equilibrium profile
(rather than an arbitrary shape for the surface density, as is 
permissible for linear analyses), and because such 1D
equilibria do not exist for very small $\Qsp$, our 2D simulations
do not cover the parts of parameter space in which $\Qsp$ is small
studied by B88.

Our hydrodynamic models are by no means exhaustive, and the uncertainties
in defining a realistic background model leave open many possibilities.
Nevertheless, both the models shown here and others we have studied
support the conclusion that without magnetic effects, it is difficult
to form large-scale spur-like structures in a gaseous arm 
that is not violently unstable to quasi-axisymmetric modes, and 
hence unlikely to represent a realistic equilibrium in the first place.

\subsection{MHD Simulations with Sub-equipartition Magnetic Fields}

For simulations with mean magnetic field strength below thermal equipartition,
we select three $\beta_0=10$ (i.e., $\vA=\cs/\sqrt{10}$)
models that are stable to quasi-axisymmetric perturbations.
The critical $Q_0$ values for $\beta_0=10$ are 1.58, 1.78, and 1.89,
for $F=1$\%, 2\%, and 3\%, respectively.
Figure \ref{MHD10evol} shows
the initial density distributions
and the evolution of maximum surface densities for
model MS1$-$MS3.

Compared to Figure \ref{HDevol}, Figure \ref{MHD10evol} 
demonstrates magnetic destabilization of modes with
$\mathbf{k}$ along the arm: perturbations in all three models 
grow very rapidly, leading to the formation of spurs 
that will eventually fragment and collapse.
This destabilization in spiral arms may be understood 
in relationship to the nonaxisymmetric
instabilities present in shearing, featureless disks.
In addition to swing amplification,
for which magnetic fields play a stabilizing role, 
magnetized thin disks are also 
subject to magneto-Jeans instability (MJI), in which 
tension forces from embedded magnetic fields resist the
stabilizing Coriolis force \citep{elm87a,elm94,kim01}.
The efficiency of the MJI depends on the local field strength and 
the local shear rate.
Even with weak magnetic fields,
significant growth of perturbations can occur, provided that
their azimuthal wavelength is less than the local Jeans
wavelength and the local shear rate is not too large
(see, e.g., Fig.\ 1 of \citealt{kim01}).
Although swing amplification may also contribute at a lower level,
background spiral-arm conditions in models MS2 and MS3 
provide fertile ground for growth of the MJI, such that the models
evolve quickly into self-gravitational runaway, within 2 orbits.
The evolution of model MS1 is similar to that of H1, in which
perturbations grow primarily via swing amplification,
although magnetic fields and stronger gravity expedite
the instability in MS1.

In Figure \ref{MHD10pat}$a$, we plot a snapshot 
of model MS3 at $t/\torb=1.78$ in the frame comoving with
the spiral pattern (that is, the frame of the simulation). 
Surface density in logarithmic color scale shows the
structure of spurs forming almost perpendicularly to
the shock, with an average spacing of $\sim0.8\,\kpc$.
The velocity field clearly shows the background shear 
of galactic rotation across the box, the streaming motion of fluid
elements within the arm, and the reversal of shear in the region of 
high density compression.
Since it amounts to only $\sim 5\%$ of the amplitude of background velocity,
however, the perturbed velocity associated with spur formation is 
difficult to see in this representation. 
The spurs move in the $y$-direction with $v_y\sim 0.50R_0\Omega_0$
with respect to the global spiral pattern, which implies 
that they follow the background galactic rotation very closely
(since $v_{\rm inertial} = R_0\Omega_p + v_y \rightarrow 0.5R_0\Omega_0
+v_y$). To see the shape of spurs clearly, 
we transform to a frame in which spurs remain 
stationary and the left $x$-boundary corresponds to the 
initial shock front location.
The evolved density structure (color scale) and streamlines (dotted lines)
in this frame are shown in Figure \ref{MHD10pat}$b$.
We also show the wavefront formed by spurs (solid line)
drawn from equation (\ref{Wfront}),
with $\kx(0)/\ky=2.5$ chosen to match the shape of spurs.
Note that except in the region very close to the shock front,
the streamlines in the stationary-spur frame follow the kinematic 
wavefronts of spurs fairly well.

Figure \ref{MHD10image} displays snapshots of density and
field configurations of models MS1 at $t/\torb=4.0$ and 
MS2 at $t/\torb=2.1$. The late-stage evolution of MS3 is similar to that 
of MS2. Our simulation resolution is insufficient
to follow the evolution of these models further.
When spurs reach high enough density, they begin to fragment along
their length. 
Each of these fragments has an average mass of
about 2.0\% and 1.5\% of the total for models MS1 and MS2, respectively,
which correspond to $\sim4\times 10^6\Msun$.
Notice that this clump mass
is roughly equal to the local Jeans mass, $M_{\rm J,sp}$, at the spiral
arm density peak
\begin{equation}\label{M_Jsp}
M_{\rm J,sp} \equiv \frac{\cs^4}{G^2\Smax}
= 7\times 10^6\Msun
\frac{\Qsp^2}{Q_0} 
\left(\frac{\cs}{7.0 {\rm\,km\,s^{-1}}} \right)^3
     \left(\frac{\kappa_0}{36\,{\rm km\,s^{-1}\,kpc^{-1}}}\right)^{-1}.
\end{equation}
From Table \ref{tbl-1}, one can see that 
$M_{\rm J,sp}$ ranges $\sim(2.5-4)\times 10^6\Msun$
for our magnetized models. Since the spur spacing is 
$\sim2-5\ljsp$, and the surface density has large gradients
across the arm width of $\sim1-2\ljsp$, one should however
not think of the clumps as being gathered isotropically 
from uniform density regions.

Some fragmentation occurs within the spiral arm
(and fragments subsequently climb out of the potential well), 
while some fragments that form off the arm are rapidly carried away deep into
the interarm region. Although the mass conversion efficiency in 
clustered star formation is uncertain, the collapsing 
fragments of our models may represent the entities that 
develop into prominent \ion{H}{2} regions, as observed in both
arm and interarm locations in spiral galaxies. 
Note that as Table \ref{tbl-1} lists, the spacing along the
arm of spurs in models MS2 and MS3 are about $\sim3-4$ times the arm width,
which is consistent with observed separation of \ion{H}{2} regions
and star complexes in spiral galaxies \citep{elm83}.

\subsection{MHD Simulations with Equipartition Magnetic Fields}

Figure \ref{MHD1evol} shows
the initial density distributions
and the evolution of maximum surface densities for
MHD simulations with $\beta_0=1$ (i.e., $\vA=\cs$).
Since magnetic fields exert pressure as if they were
adiabatic gas with an ``effective'' index
of two for one-dimensional compression 
(cf.\ \citealt{shu92}), 
they potentially can reduce the density compression greatly 
even in an isothermal shock.
The maximum densities in the resulting steady-state shock profiles 
in models ME1$-$ME4 are less than 3, and the maximum to 
minimum density ratios are smaller than 6 as well.
The minimum value of $\beta$ is $\sim0.3-0.5$.
With the corresponding values $\Qc \sim 0.86-0.89$ for 
the existence of equilibrium configurations,
the chosen models having $Q_0=1.2-1.6$ are quite
far from the margins of susceptibility for quasi-axisymmetric 
instability. Nevertheless, these $\beta_0=1$ models respond with
strong amplification of perturbations that vary along
the spiral arm, forming self-gravitating
structures within $\sim2.5$ orbits, as seen Figure \ref{MHD1evol}$b$.

In order to study the morphology and kinematics of spurs,
we display in Figure \ref{MHD1pat} various aspects of the surface 
density and velocity 
structure of model ME3 at $t/\torb=1.27$.
Figure \ref{MHD1pat}$a$ clearly shows that nine spurs, 
whose density is displayed in logarithmic color scale, grow 
out of the spiral arm.
Velocity vectors measured in the spiral arm frame
show the background shear, reversed shear, and
streaming motions associated with passage through the spiral arm, 
but as for model MS3 (Fig.\ \ref{MHD10pat}$a$) they do not 
readily evidence the perturbed velocity components that create spurs. 
The spurs of model ME3 have an average spacing of $\lambda_y\sim 0.7\kpc$,
and move along the arm with $v_y\sim 0.47R_0\Omega_0$. 
Figure \ref{MHD1pat}$b$ shows that 
the shape of the spurs is in excellent agreement with the 
kinematic prediction for wavefronts (solid lines) 
based on background shear and expansion
from equation (\ref{Wfront}) with $\kx(0)/\ky=0.2.$\footnote{With
$\kx(0)/\ky>0$, the spurs are trailing when they leave the
shock front ($\tau=0$). Since $\kx(0)/\ky=0.2$ is relatively small, 
however, any appreciable $\tau$ in eq.\ (\ref{Wfront}) 
gives $\kx/\ky=-\mathcal{T}<0$ so
that wavefronts become leading almost immediately downstream of the 
shock front, as Fig.\ \ref{MHD1pat}$b$ shows.}
Convergence of streamlines (dotted lines) towards the regions of
high density in spurs suggests that they grow
by accumulating material mostly in the $y$-direction,
which in turn implies that spurs in model ME3 are products of
MJIs \citep{elm87a,kim01}.

To provide a feeling for spur-region kinematics as it could 
be observed with a radio interferometer or Fabry-Perot 
instrument, we create sample synthetic maps. 
Imagine a spiral galaxy whose local disk can be identified with
model ME3 at $t/\torb=1.27$. The galaxy is assumed to
be inclined arbitrarily by 50 degrees with respect to the plane of sky.
For fictitious observations, we select two viewing angles:
parallel to the spiral arm and 40 degrees off the arm.
In both cases, the target areas are rotating away from 
the observer. The resulting line-of-sight velocities (color scale)
and ``deprojected'' column density (contours) are presented, 
respectively, in Figures \ref{MHD1pat}$c$ and \ref{MHD1pat}$d$. 
The numbers in the color bars are in the units of km s$^{-1}$.
When viewed along the spiral arm,
the line-of-sight velocity is equal to $v_y\sin\alpha$
(where $\alpha$ is the angle of the disk plane with respect to
the plane-of-sky).
With this observer position angle, the signature of
nonuniform streaming motions influenced by the external potential is evident,
as manifested by the small change in color gradient across the arm
in Figure \ref{MHD1pat}$c$. 
The line-of-sight velocity has no discontinuities in this case.
When the line-of-sight is oblique to the arm,
on the other hand, the line-of-sight velocity is a mixture 
of the velocities both parallel and perpendicular to the arm.
Correspondingly, the signature of parallel
streaming motion in the line-of-sight velocity
is suppressed relative to the signature of the large discontinuity 
in $v_x$ across the shock, as shown in Figure \ref{MHD1pat}$d$ . 
A careful examination of Figures \ref{MHD1pat}$c$ and \ref{MHD1pat}$d$
reveals that the line-of-sight velocity is correlated with
spurs, with high-density regions having negative gradients in the
velocity ($dv_y/dy<0$), indicative of converging flow. 
The amplitude of the variation in the line-of-sight velocity along the arm 
is typically $\sim 5$ km s$^{-1}$ in the spur regions, 
which is almost comparable 
to the variation of streaming motion (i.e., shear in $x$ of $v_y$)
in the arm.

As mentioned before, the primary mechanism for structure formation 
in $\beta_0=1$ models is the MJI, in which threaded near-azimuthal
magnetic fields transport angular momentum out of growing perturbations.
Time evolution of the linear MJI in a disk
with uniform density, uniform magnetic fields, and normal linear
shear has been studied by \citet{elm87a}, \citet{gam96},
\citet{fan97}, and very recently by \citet{kim01} -- in which
nonlinear evolution was also studied.
In previous works, however, the effects of varying shear and 
background compression and expansion across a spiral shock were not
taken into account.
Because of the evidence from our simulations of the importance of MJIs,
it is valuable to generalize the analytical work of B88 on
hydrodynamical self-gravitating instabilities in spiral arms
to include magnetic effects. This will enable us to make direct 
comparisons between the linear evolution of MJIs 
in non-uniform media
and the results of our numerical simulations.

In Appendix A, we present the linearly perturbed MHD equations 
in the Lagrangian frame moving with the unperturbed 
background flow through a spiral arm.
In order to integrate the governing set of equations,
we need to prescribe an equilibrium density configuration.
For an exemplary run, we adopt that of model ME3
(with $Q_0=1.5$, $\beta_0=1$, and $F=3\%$) as shown in Figure \ref{1Dshock1}.
Just as in B88, we follow the growth of a disturbance assumed to maintain
phases of an initial locally plane-wave form 
$\propto e^{i(\kx(0) x_0 + \ky y_0)}$,
where $x_0$ and $y_0$ are the initial position of a fluid element.
We are free to choose $\kx(0)/\ky$ and $K_y$ of any input perturbations.
Taking $\delta\Sigma/\Sigma\neq0$ and $\dvx=\dvy=dm=0$ as initial conditions
at the shock front, and varying $\kx(0)/\ky$ and $K_y$,
we directly integrate equations (\ref{ppcon})$-$(\ref{ppind}) over time.
For a given wavelet, we measure the total amplification factor 
and the time $\tau_{\rm grow}$ when maximum amplification occurs, 
and plot these values in Figure \ref{mhd_con} as solid and dotted contours,
respectively.
Overall, modes with $K_y<0.5$ are favored for growth, 
in sharp contrast to MJIs in a uniform medium where dominant modes 
have $K_y \sim \case{1}{2}-\case{3}{4}$.
This is because $\kjsp$ used in the definition of $K_y$ 
is in fact the maximum possible value in the medium.  
The value of $k_{\rm J} = 2\pi G\Sigma/\cs^2$ averaged over the arm 
is lower, rendering the ``average'' value of $\ky/k_{\rm J}$ larger
than $K_y \equiv \ky/\kjsp$.

Since MJIs generally require the wavenumber of perturbations to be less
than the local Jeans wavenumber for instantaneous growth,
the total amplification factor depends on how long 
the kinematics of the background flow can keep $\kx$ small
such that $(\kx^2+\ky^2)^{1/2}$ is smaller than the local $k_{\rm J}$
(which itself decreases as the surface density drops outside
the spiral arm).
While $\kx$ is a linearly increasing function of time in a 
normal, uniform-shear case, 
the evolution of $\kx$ in the spiral arm is not generally monotonic,
determined instead by equation (\ref{Wfront}). 
It turns out, as Figure \ref{mhd_con} shows,
that the MJIs associated with the spiral arm 
prefer wavelets that are initially slightly trailing at the shock front
($\kx(0)/\ky>0$). This is because
the shear reversal inside the spiral arm produces an interval in which
$\kx$ initially decreases, followed by the return of normal shear
downstream which then increases $\kx$. The expansion effect also
tends to reduce $\kx$ if it is initially positive.
Starting a wavelet as trailing thereby extends the time span 
during which MJIs operate efficiently, and thus yields maximum amplification.

In Figure \ref{mhd_con}, we also indicate the parameters 
($K_y=0.45$ and $\kx(0)/\ky=0.2$) 
determined from the numerical simulation for model ME3 (see
Fig.\ \ref{MHD1pat}) as a rectangular box. 
The ambiguities in determining $\ky$ from an {\it integral} number of 
spurs and in finding $\kx(0)/\ky$ that gives the best fit for the shape
of spurs are represented by the box size.
The corresponding amplification factor and growth time from
the linear-theory integration are
$\sim60-80$ and $\tau_{\rm grow}\sim3.1-3.5$, respectively. 
While the largest predicted linear amplification available for the system 
is attained at the same $\kx(0)/\ky=0.2$, but at a larger 
length scale with $K_y=0.32$,
this mode (having $\tau_{\rm grow}=3.6$) would take longer
to achieve the maximum growth.
It appears that the nonlinear system ``compromises'' between
maximum amplification and earliest growth, in selecting the mode that
dominates the evolved state of the simulation. 

In Figure \ref{bc_comp}, we compare the variation in the 
perturbed surface 
density distribution within the 
spurs with the prediction of amplification from the linear analysis.
In doing this, we use the equivalence between the spatial variable $x$
in the simulation and the Lagrangian distance traveled downstream
by a fluid element in the background state, in the linear analysis.
Linear evolution of the perturbed surface density with $K_y=0.45$
and $\kx(0)/\ky=0.2$ is plotted as a function of the dimensionless
Lagrangian time variable $\tau$, and the corresponding downstream 
distance from the shock front. 
Note that $\tau$ varies from 0 to $2\pi m^{-1}(1-\Omega_p/\Omega_0)^{-1}$ 
for $x/L_x=0-1$, corresponding to $\tau=0-2\pi$ for two-armed spirals
with $\Omega_p=\Omega_0/2$.
We select three prominent spurs in Figure \ref{MHD1pat}$b$ and
plot with various curves 
their perturbed surface densities measured along their length.
Although densities near the shock front fluctuate with small
amplitudes due to the interactions with spurs' trailing tails,
the overall density distribution and the positions where spurs
achieve maximum densities are in remarkably good agreement with the results
of linear analysis.

The growth of perturbations depends sensitively on $Q_0$.
For instance, model ME4 with $Q_0=1.6$ takes about one more
orbital time than model ME3 with $Q_0=1.5$ to be fully
nonlinear, although initial density distributions are
quite similar to each other (see Fig.\ \ref{MHD1evol}).
This is not because the linear analysis for $Q_0=1.6$
predicts a larger value of $\tau_{\rm grow}$ than for
$Q_0=1.5$, but because the perturbations keep
growing as they pass successively through spiral arms.
As $Q_0$ increases further, linear amplification factors
become smaller, so that higher-$Q_0$ models need 
even longer time for development. We found that a model
with $Q_0=1.7, \beta_0=1$, and $F=3\%$ (not listed in
Table \ref{tbl-1}) becomes only
moderately nonlinear at $t/\torb=4$,
while models with $Q_0\geq 1.8$ and
the same $\beta_0$ and $F$ remain linear until the end
of simulations ($t/\torb=5$).

Finally, Figure \ref{MHD1image} displays developing and fragmenting 
spurs for models ME3 at $t/\torb = 1.59$ 
and ME4 at $t/\torb = 2.55$. 
The separation between two neighboring spurs is about
$\sim 0.7-0.8$ kpc, which is very close to the  most
unstable Jeans wavelength at the spiral density peak
($\Kymax\sim0.45$; see Table \ref{tbl-1}),
showing that the stabilizing contribution from epicyclic motions
is almost negligible when the magnetic field is strong.
Compared to $\beta_0=10$ models, spurs in the $\beta_0=1$ models 
are stronger and extend farther away from the spiral arm. 
Fragmentation of spurs when the magnetic field is stronger thus 
occurs mostly in the interarm region.
For these runs, we find that
each collapsing fragment has an average mass of
about 1.5\% ($\sim 4\times 10^6\Msun$) of the total mass,
which is again very close to the local Jeans mass
$M_{\rm J,sp}$ at the peak density of spiral arms (see eq.\ [\ref{M_Jsp}]
and Table \ref{tbl-1}).
One cannot resist speculating that bright, interarm \ion{H}{2}
regions, as seen for example in the optical image of M51, could
have originated as condensations in spur structures very like
those found by our simulations and analyses.

\section{Discussion}

\subsection{Summary of Modeling and Results}

In this paper, we have investigated the dynamical interaction of
a magnetized, 2D shearing flow -- representing a local
patch of the height-integrated ISM -- with an external 
gravitational potential -- representing a local portion 
of a spiral stellar arm.
Our primary interest was in exploring the ways in which self-gravity
leads to growth of intermediate-scale structure, and how
these intermediate-scale structures subsequently fragment.
We show that in magnetized systems, the characteristic 
structures that grow are spur-like features that jut out at
regular intervals from the spiral arms;
within these spurs, dense condensations may grow and
travel out into the interarm region. The spur structures
in our models bear remarkable resemblance to the 
conspicuous features branching out of spiral arms in recent
high-resolution HST images of the Whirlpool galaxy, and 
also apparent (less spectacularly) in other grand-design spirals.

The technical centerpiece of our work is a set of 
time-dependent, numerical MHD simulations. 
The model disks are infinitesimally thin and maintain a constant 
isothermal sound speed $\cs$.  
In the absence of stellar spiral-arm forcing, 
the disks are assumed to have uniform velocity shear, 
uniform surface density characterized
by the Toomre $Q_0$ parameter (see eq.\ [\ref{Q0}]), and uniform
near-azimuthal magnetic fields, with $\beta_0\equiv\cs^2/\vA^2$
measuring the mean field strength at the disk midplane
(see eq.\ [\ref{beta}]). We do not allow for spacetime variations
of the magnetic scale height.
In addition to its own self-gravity, the gaseous flow is subject to
an explicitly-introduced sinusoidal external 
potential (see eqs.\ [\ref{ext_P}] and [\ref{F_arm}]),
modeling a local stellar spiral arm.
The back reaction of stars to the gravitational field of gas
is not, however, included in the present study.
The spiral arm is assumed to be tightly wound and rigidly rotating
at half the local orbital rate.
We evolve the ideal MHD equations in the
local frame comoving with the stellar arm, with two orthogonal axes 
corresponding to the directions perpendicular and parallel to the arm
(see \citealt{rob69}).

Imposing translational symmetry along the arm, 
we first construct one-dimensional equilibrium spiral shock 
configurations
and examine their stability to axisymmetric perturbations.
Even if the spiral perturbation is only a tiny fraction of the background
axisymmetric field, induced radial velocities are generally supersonic,
easily forming a shock front.
We find that larger amplitude background spiral arms (larger $F$)
result in stronger shocks,
while the shock-forming tendency of gas decreases as the relative
importance of self-gravity to spiral forcing increases 
(cf.\ \citealt{lub86}).
Self-gravity enhances the arm-interarm
density contrast, while magnetic pressure reduces it.
When $Q_0$ is sufficiently small, no equilibrium profiles 
can be found from our time-dependent integrations,
because either the structures become gravitationally unstable (when
magnetic fields and/or stellar spiral perturbations are weak), 
or because the given background conditions and adopted spiral pattern
speed do not allow stationary shocks to exist (when fields and/or
spiral arms are strong). Critical values of $Q$ at the density
peak are found to be $\Qsp\thickapprox 0.8,$ 0.5, 
and 0.4 for $\beta_0=\infty$, 10, and 1, respectively.

For our two-dimensional simulations, we start with
one-dimensional equilibrium density profiles calculated as above,
apply low-amplitude white-noise perturbations, 
and monitor their nonlinear growth.
We find that gaseous spurs naturally form as a consequence of
gravitational instabilities inside spiral arms with
input modes having $\mathbf{k}$ nearly along the spiral arms
growing preferentially.
The chief physical mechanism for spur formation is
the magneto-Jeans instability, in which
the magnetic tension force breaks the loop of stabilizing epicyclic
motions \citep{elm87a,kim01}. 
Although swing amplification can also play a role
in forming spur structures when the magnetic field is weak and
the arm-interarm density contrast is moderate,
purely hydrodynamic systems that are stable to quasi-axisymmetric
perturbations are generally also found to be stable to nonaxisymmetric
perturbations. We thus suggest that magnetic effects are key for 
producing the gaseous spurs observed in real galaxies.

When the mean magnetic field is in thermal equipartition ($\beta_0=1$),
we find that the separation of spurs is more-or-less
consistent with the most unstable Jeans wavenumber at the density peak,
while with a sub-equipartition magnetic field ($\beta_0=10$),
the formation of spurs requires larger-scale perturbations to
enhance self-gravity (see Table \ref{tbl-1}).
The characteristic scale of spurs formed in our simulations is
about $\lambda_y\sim 2.5\ljsp$ (or $\sim 750$ pc for our model
parameters) for the models that are unstable to the magneto-Jeans 
instability, where $\ljsp$ is the local Jeans wavelength at the 
spiral-arm density peak. 
The corresponding ratio of the spacing to the arm thickness
is $\lambda_y/W\sim 3.5$ and 1.5, respectively, for $\beta_0=10$ and 1 models.
These ratios are consistent with the observational 
results for the distribution of \ion{H}{2} regions along spiral arms
\citep{elm83}. 

The shape of the spurs sculpted by the shearing and expanding background 
flow in our models is in excellent agreement with 
equation (\ref{Wfront}), originally derived by B88.
Reversed shear within the strongly compressed region tends to decrease the 
$x$-wavenumber of the perturbations, so that
slightly trailing waves entering the spiral shock 
are preferred as having the longest growing phase under MJIs.
Entering waves first shear around and expand to create nearly-radial
wavefronts as they exit the arm.
Far into the interarm region, where shear returns to the ``normal'' 
(slower-outside) sense, the wavefronts become strongly trailing.

In the nonlinear stage of evolution, spurs experience fragmentation to
form gravitationally bound clumps, the typical mass of which
corresponds roughly to
the local Jeans mass at the density peak inside spiral arms
(or $\sim4\times 10^6\,\Msun$ for our model parameters).
The positions where the fragmentation
occurs include interarm regions as well as within spiral arms themselves.
We propose that interarm condensations formed in this way could 
evolve into bright \ion{H}{2} regions in interarm regions.

\subsection{Application to Spiral Galaxies}

Now we apply our simulation results to a grand-design spiral, M51.
We are particularly interested in a section of the southern spiral arm
at 2.7 kpc from the center, where 
spurs can be relatively easily identified and observed data are available. 
Here, we adopted a galactic distance of
9.6 Mpc to M51 \citep{san75}. From the
recent combined WFPC2/NOAO Hubble Heritage image of M51 
\citep{sco01}, we recognize five spurs in this portion of the arm
with an average spacing of $\lambda_y \sim 350$ pc. 
From radio interferometry and single dish observations, 
$\Omega_0\sim 78 {\,\rm km\,s^{-1}\,kpc^{-1}}$,
$\Sigma_0 = 80  \,\Msun\pc^{-2}$, and 
$\Sigma_{\rm max} = 170  \,\Msun\pc^{-2}$ \citep{ran93,gar93}. 
With $\cs\approx 10{\,\rm km\,s^{-1}}$, 
the values listed above give $\ljsp \sim 140$ pc,
corresponding to $\Kymax\equiv \ljsp/\lambda_y\sim 0.4$.
This is consistent with Table \ref{tbl-1},
which suggests that spurs produced by MJIs have 
$\Kymax\sim 0.3-0.5$. 

Giant \ion{H}{2} regions are well distributed along
the dominant spiral arms in grand design spirals, 
but a significant fraction of them are also found in 
deep interarm regions. \citet{sco01} reported that about 55\% of 
identified \ion{H}{2} regions are in interarm regions, although
overlapping of \ion{H}{2} regions in the HST image of M51 
could reduce their numbers in the arms. 
Given the nominally low surface density and high shear
conditions in interarm regions, the presence of numerous interarm
\ion{H}{2} regions requires an explanation. From our model simulations,
we suggest that massive self-gravitating clumps resulting from nonlinear 
fragmentation of spurs can develop in both arm and interarm regions.
This implies that interarm star formation may be engendered by
the same dynamical process, namely large-scale gravitational
instabilities initiated in spiral arms, as arm star formation. 
Indeed, there is no significant
observational difference in the sizes and electron densities
of the \ion{H}{2} regions between spiral arm and interarm regions
\citep{sco01}.

The fact that the shape of spurs, reflecting the background flow
characteristics, is well described by equation (\ref{Wfront})
may provide an independent way to determine global spiral pattern speeds.
The pitch angle, angular velocity of galactic rotation,
and epicyclic frequency are observable quantities, and 
$\ky$ can be determined from the mean separation of observed 
spurs. Provided that the surface density distribution is known,
therefore, $\kx(0)$ and $\Omega_p$ (see eq.\ [\ref{tele}])
can be determined simultaneously by
finding a best fit of $\mathcal{T}$ to the observed overall shape of spurs. 
If we can further constrain $\kx(0)/\ky$ for example from 
the linear theory as  the dominant mode, or from
high resolution observations
that allow determination of the slope of spurs
near the density maximum ($\tau\rightarrow0$), 
then $\Omega_p$ can be uniquely determined.
Conversely, if we know the pattern speed by other means, 
equations (\ref{Wfront}) and (\ref{tele}) can be used 
to predict the variation of surface density in spiral arms,
as a consistency check of direct determinations.

For model simulations presented in this paper, we 
took a patch of a galactic disk inside the corotation 
radius of a steady spiral pattern, and showed that
spurs jut outwards from the arms and become increasingly trailing
at larger radii.
We have also run spiral-arm interaction simulations for regions
outside the pattern's corotation (not listed in Table \ref{tbl-1})
and found that the induced spurs jut {\it inwards}
from the local arm segments and become trailing at smaller radii.
These morphological differences in spurs between inside and outside 
corotation could help to locate corotation radii of 
spiral patterns in spiral galaxies. 
The present lack of evidence for spurs on the insides of arms 
(B.\ Elmegreen, personal communication)
could be because the arms are generally relatively weak
outside corotation, or perhaps because resolution is insufficient.  
Of course,
it is possible that when other galaxies are observed at resolution
comparable to the recent M51 image, and when larger areas are
surveyed, inward-jutting spurs might be detected.

\subsection{Outstanding Issues}

In this paper, the 2D simulations of purely hydrodynamic flow
passing through spiral arms show comparatively stable behavior,
largely because the hydrodynamic models chosen for these runs 
have relatively large $\Qsp>0.9$ at the density peak. 
\citet{bal85} argued that for flows with pressure-density 
relation $\Pi\propto \Sigma^\gamma$, the values of the 
Toomre parameter at the point of maximum compression,
$\Qsp\equiv Q_0\gamma^{1/2}
(\Smax/\Sigma_0)^{\gamma/2-1}$, is the most
important parameter for characterizing the system's gravitational
response (here $\Sigma_0$ and $Q_0$ are the equivalent parameters 
for a uniform disk). Since $\Qsp$ is the smallest local value in a non-uniform
medium, $\Qsp<1$ does not always ensure quasi-axisymmetric instability
will occur, because for local instability $Q$ must be smaller than
unity over at least a radial distance $\sim\lambda_{\rm max}
= 2\cs^2/G\Sigma = 2\gamma c_{s,0}^2 (G\Sigma_0)^{-1}
(\Sigma/\Sigma_0)^{\gamma-2}$. 
Our results for quasi-axisymmetric stability indicate 
that critical $\Qsp$ values are $\sim 0.73-0.98$ for unmagnetized
cases (see \S3 and Fig.\ \ref{Qcrit}).
It seems that at least one additional parameter, $W$, 
which describes arm
width (or the gradient in the background density profile), combines
with $\Qsp$ to determine the system's response to two-dimensional
(spur-forming)
perturbations. For example, models H2 and H3 having smaller
$\Qsp$ remain stable (with fluctuating density),
while model H1 with larger $\Qsp$ (but smaller $Q_0$)
is subject to (mild) swing amplification.
For a given $\Qsp$, models with larger $W$ -- and thus a 
smaller arm-to-interarm density contrast and $Q_0$ -- 
are usually more susceptible to swing amplification.
In part, this is because excessive compression in the spiral arm
reverses the normal direction of shear (see eq.\ [\ref{loc_shear}]),
which tends to suppress the swing mechanism.
For wavelets that enter the spiral shock with leading 
orientation ($\kx <0$), B88 showed that the tendency for ``normal''
rotation of leading wavefronts driven by expansion
(see eq.\ [3.15] of B88) can compensate somewhat for locally-reversed shear.
We find that for cases where quasi-axisymmetric modes are stable,
however, amplification from ``natural'' perturbations (mostly 
trailing) does not produce strong spurs in unmagnetized systems.

From B88's results with $\gamma=0.5$, 
$\Qsp<0.7$ appears to be required 
in order for {\it nonaxisymmetric}
perturbations to grow significantly enough to form nonlinear structures.
With a small value of $\gamma$, moderate compression factors 
easily produce $\Qsp$ as low as 0.6 from mean ISM conditions.
However, {\it quasi-axisymmetric} modes may be unstable as well when
$\Qsp$ is very small.
By adopting an isothermal equation of state,
we obtain larger values of $\Qsp$ for a given compression factor
than with a softer pressure-density relation.
In addition, because our 2D models are initiated with 1D profiles 
that are quasi-axisymmetrically stable,
we were limited to a range of $\Qsp>0.9$ for our hydrodynamic models.
Although introduced to some extent to achieve numerical stability
in our computations, the requirement of quasi-axisymmetric stability
is likely to appear in nature as well.
Since we observe gaseous spiral arms and associated dust lanes in
spiral galaxies,
the compressed gaseous regions are probably quasi-axisymmetrically
stable; otherwise, these arms would not be long-lived -- 
compressed gas would turn quickly into stars. 

The question, then, of 
whether spurs could form from purely hydrodynamic (swing-like) 
instabilities within arms is reduced to finding a self-consistent
density distribution that is quasi-axisymmetrically stable
yet with $\Qsp$ small enough for 2D instabilities to grow. 
The real ISM is multiphase and turbulent,
in which hot, warm, and cold 
phases coexist (e.g., \citealt{fie69,cox74,mck77,hei01}). 
Although optically thin gas can be described by
polytropic relations between density and thermal pressure,
this is only in a piecewise sense with a fairly uncertain $\gamma$
(see, e.g., \citealt{vaz00}).
The turbulent motions of cold clouds, although often treated
as an effective pressure, may actually have a rather
different dynamical response to compression.
A realistic assessment of hydrodynamic instabilities on
spur and cloud formation, therefore, will require a more accurate
treatment of the small-scale thermal and dynamical properties
of the medium. An important direction for future research, thus,
will focus on determining whether better treatment of 
``microphysics'' renders the purely-hydrodynamic (swing)
mechanism competitive with the MHD mechanism (MJI) for
forming spurs and condensing clouds.

In the present paper, we have studied the formation and fragmentation 
of spurs due to self-gravitating instabilities within spiral arms,
with magnetic fields parallel to the galactic plane abetting the process. 
Since they are restricted to a thin-disk geometry, the present models do not
capture the potential dynamical consequences of magnetorotational instabilities
(MRIs) and the Parker instability in themselves or in connection to MJIs.
MRIs (e.g., \citealt{bal98}), which exist only in 3D systems, are known
to generate MHD turbulence, which may make an important contribution
to the amplitude of ISM random motions in galactic disks \citep{sel99}
as well as exciting density perturbations, and may also provide angular
momentum transport within disks -- possibly affecting the growth of
condensations.

The \citet{par66} instability 
has long been thought to be important in the formation 
of OB associations, giant \ion{H}{2} regions, and 
giant molecular clouds along spiral arms (e.g., \citealt{mou74,bli80}),
because it is able to grow condensations at lateral wavelengths
comparable to the disk scale height -- as required for clouds of mass
$\sim 10^5 \Msun$.
Recent numerical studies indicate that 
the Parker instability {\it alone} cannot be the main formation mechanism 
for giant clouds in general galactic disks,
because the column density enhancement 
is only a factor of three (\citealt{kimj00,san00}; see also
\citealt{elm95}).
However, the coupling of the Parker instability with other dynamical
processes could produce significant effects on structure formation
in galaxies.
\citet{bas97} suggested that
nonlinear triggering of the Parker instability by spiral
shocks could enhance density enough to form giant cloud complexes,
and the nonlinear simulations of \citet{cho00} (neglecting shear)
evidence this sort of behavior.
Extension of the current work
into three dimensions has the potential to show how ``seeds'' sown
by the Parker and/or MRI modes
in spiral arms may grow and condense into self-gravitating
giant molecular clouds.

\acknowledgements

It is a pleasure to acknowledge stimulating conversations and
communications with B.\ Elmegreen,
J.\ Stone, S.\ Vogel, F.\ Shu, N.\ Scoville,
and K.\ Sheth. We are also grateful to an anonymous referee for
constructive comments.
This work was supported by NASA grants NAG 53840 and NAG 59167.

\appendix
\section{Linear Analysis}

In this appendix, we provide the perturbed equations to be 
integrated to obtain the linear-theory solutions shown in 
Figures \ref{mhd_con} and \ref{bc_comp}.
These solutions represent examples of
the evolution of low-amplitude, nonaxisymmetric perturbations
(specifically, initial local plane waves) 
in a magnetized, non-uniform disk, subject to a background flow containing
compression and expansion, and varying shear, as imposed by
the self-consistent equilibrium response to an
external spiral potential.
Local dynamical instabilities in {\it unmagnetized}
disks with similar background flow properties have been
studied by \citet{bal85} for quasi-axisymmetric modes 
and by B88 for two-dimensional perturbations.
Here, we extend B88's work to include the effect of 
magnetic fields.

We begin by considering a steady-state equilibrium 
configuration represented by density $\Sigma$, velocity 
$\mathbf{v}=u\mathbf{\hat{x}} + v\mathbf{\hat{y}}$,
and magnetic field $\mathbf{B} = B_y \mathbf{\hat{y}}$ (see for
example \S 3), in the spiral-arm coordinates built in \S 2.1.
The equilibrium is quasi-axisymmetric in the sense that
flow quantities vary only along $\mathbf{\hat{x}}$, the
coordinate axis perpendicular to the arm, which lies
at an angle $i$ with respect to the local radial direction.
Imposing small amplitude Eulerian perturbations represented by $\delta$,
we linearize equations (\ref{con})-(\ref{eos}):
\begin{equation}\label{pcon}
\frac{d}{dt}\left(\frac{\delta\Sigma}{\Sigma}\right) = 
-\frac{\partial\delta u}{\partial x} 
-\frac{\partial\delta v}{\partial y} 
-\delta u \frac{d\ln{\Sigma}}{dx},
\end{equation}
\begin{equation}\label{pmomx}
\frac{d\delta u}{dt}  
= - \delta u \frac{du}{dx} + 
2\Omega_0 \delta v 
-\frac{\partial}{\partial x} \left(\cs^2\frac{\delta \Sigma}{\Sigma} 
+ \delta\Phi_g\right)
+ \frac{B_y}{4\pi\Sigma}
 \nabla^2
 \delta m
+\frac{1}{4\pi \Sigma}\frac{dB_y}{dx}
     \left(\frac{\partial\delta m}{\partial x} + 
      B_y\frac{\delta\Sigma}{\Sigma}\right),
\end{equation}
\begin{equation}\label{pmomy}
\frac{d\delta v}{dt}  =
-\left(\frac{\kappa_0^2}{2\Omega_0} + \frac{dv}{dx} \right)\delta u
-\frac{\partial}{\partial y}
\left(\cs^2\frac{\delta \Sigma}{\Sigma}+ \delta\Phi_g\right)
+ \frac{1}{4\pi\Sigma}\frac{dB_y}{dx}\frac{\partial\delta m}{\partial y},
\end{equation}
\begin{equation}\label{pind}
\frac{d\delta m}{dt} = B_y\delta u,
\end{equation}
\begin{equation}\label{ppos}
\nabla^2
\delta\Phi_g
= 4\pi G \delta(z) \delta\Sigma,
\end{equation}
where
the Lagrangian time derivative following the background flow
is denoted by
\begin{equation}
\frac{d}{dt} \equiv \frac{\partial}{\partial t}
+ u_T \frac{\partial}{\partial x} 
+ v_T \frac{\partial}{\partial y},
\end{equation}
and the perturbed vector potential
$\delta m$ is defined through 
$\delta\mathbf{B}\equiv \nabla\times(\delta m \mathbf{\hat{z}})$.

Following B88,
we now consider a Lagrangian frame
comoving with the background flow $\mathbf{v}_T$, 
allowing for the local normal or reversed shear and expansion
velocity fields.
The frame is initially located at the postshock density peak.
We adopt the ansatz that an applied plane-wave disturbance
locally preserves sinusoidal variations, so that the stability of a local 
patch can be assessed by following the temporal evolution of wavelets
of arbitrary initial local $\mathbf{k}(0)$. 
Strictly speaking, this sort of analysis should be applied only
when the initial wavelength is small compared to the width
of the spiral arm. Formally, however, we can choose $\mathbf{k}(0)$ freely.
Since the background variables are independent of $y$,
the $y$-wavenumber $\ky$ of the perturbations remains fixed 
throughout the linear evolution, 
but the $x$-wavenumber $\kx$ changes in response to the background
flow. In particular, normal/reversed shear tends to increase/decrease
$\kx$, and expansion/contraction tends to decrease/increase $|\kx|$.
The net result of these various effects on the local value of
$\kx$ was derived by B88; our equation (\ref{Wfront}) reproduces his
formula for the local value of $\kx/\ky=-\mathcal{T}$ as 
a function of the elapsed time since entering the shock (or,
correspondingly the perpendicular $x$-distance traversed).

For consistency with other ``local'' simplifications we have made,
we shall assume that spatial variations of 
the perturbed variables are much more rapid than those of the 
background state, so that we can
ignore the last term in each of equations 
(\ref{pcon})$-$(\ref{pmomy}). The validity of this approximation
can easily be demonstrated by computing 
$r\equiv \ky(d\ln{\Sigma}/dx)^{-1}$; if $r\gg1$, we can safely ignore
the terms containing the gradients of $\Sigma$ and $B_y$
in equations (\ref{pcon})$-$(\ref{pmomy}) (recall that 
$\Sigma\propto B_y$ from flux conservation). 
Since $d\ln{\Sigma}/dx\sim W^{-1}$, where $W$ is 
the arm width (cf.\ Table \ref{tbl-1}), it follows that 
\begin{displaymath}
r \sim W\ky \sim 33 \left(\frac{K_y}{Q_0}\right)
\left(\frac{\Smax}{\Sigma_0}\right)
\left(\frac{W}{L_x}\right),
\end{displaymath}
where $K_y\equiv \ky/\kjsp$, with the Jeans wavenumber $\kjsp$
at the density peak defined by equation (\ref{k_Jsp}).
Using the background parameters listed in Table \ref{tbl-1}
with the identification of $K_y=\Kymax$, corresponding to the most
unstable modes, we find $r\sim 4-5$ for the $\beta_0=1$ models
and $r\sim 1-2$ for the $\beta_0=10$ models. This implies that
the WKB approximation we make here is somewhat marginal for the 
$\beta_0=1$ models, and probably breaks down for the $\beta_0=10$ models.
For this reason, we shall apply the results of the linear analysis
only for the $\beta_0=1$ models, particularly model ME3.
 
Defining the dimensionless variables,
$\tau\equiv t\Omega_0$,
$\delta \sigma \equiv \delta \Sigma/\Sigma$,
$\delta \tilde{u} \equiv i\delta u\kjsp/\Omega_0$,
$\delta \tilde{v} \equiv i\delta v\kjsp/\Omega_0$, and
$\delta \tilde{m} \equiv i\delta m\kjsp/(\mathcal{R}B_y)$,
and applying the WKB approximation,
we now rewrite equations (\ref{pcon})-(\ref{ppos}) in 
dimensionless form (omitting the tilde) as
\begin{equation}\label{ppcon}
\frac{d\delta\sigma}{d\tau} = K_y(\mathcal{T}\delta u - \delta v),
\end{equation}
\begin{equation}
\frac{1}{\mathcal{R}}\frac{d(\mathcal{R} \delta u)}{d\tau} = 
2\delta v - \alpha K_y\mathcal{T}
\left[1 - \frac{1}{\mathcal{R}|K_y|(1+\mathcal{T}^2)^{1/2}}\right]\delta\sigma
- \alpha \beta_0^{-1} \sigma_{\rm max} K_y^2 (1+\mathcal{T}^2)\delta m,
\end{equation}
\begin{equation}
\frac{d\delta v}{d\tau} = 
-(2-q_0)\sigma_{\rm max}\frac{\delta u }{\mathcal{R}}
+ \alpha K_y 
\left[1 - \frac{1}{\mathcal{R}|K_y|(1+\mathcal{T}^2)^{1/2}}\right]\delta\sigma,
\end{equation}
\begin{equation}\label{ppind}
\frac{d\delta m}{d\tau} = \frac{\delta u}{\mathcal{R}},
\end{equation}
where $\mathcal{R}\equiv \Smax/\Sigma$, 
$\sigma_{\rm max} \equiv \Smax/\Sigma_0$,
and $\alpha \equiv (\cs\kjsp/\Omega_0)^2=8(2-q_0)\sigma_{\rm max}^2
Q_0^{-2}$. 

For temporal integrations of equations (\ref{ppcon})$-$(\ref{ppind}),
we need to specify $\mathcal{R}=\mathcal{R}(\tau)$ and $\sigma_{\rm max}$ from
a given background flow model: we adopt those of model ME3 shown in 
Figure \ref{1Dshock1}. For initial conditions, we take
$\delta \sigma=1$, and $\delta u=\delta v = \delta m =0$. 
We vary $K_y$ and $\kx(0)/\ky$ and compute amplification factors.
The results are presented in Figures \ref{mhd_con} and \ref{bc_comp}.

\clearpage
\begin{deluxetable}{cccccccccc}
\tabletypesize{\footnotesize}
\tablecaption{Parameters of Nonaxisymmetric Simulations.
\label{tbl-1}}
\tablewidth{0pt}
\tablehead{
\colhead{\begin{tabular}{c} Model\tablenotemark{a}      
                                       \\ (1) \end{tabular} } &
\colhead{\begin{tabular}{c} $Q_0$      \\ (2) \end{tabular} } &
\colhead{\begin{tabular}{c} $\beta_0$  \\ (3) \end{tabular} } &
\colhead{\begin{tabular}{c} $F$        \\ (4) \end{tabular} } &
\colhead{\begin{tabular}{c} $\Sigma_{\rm max}/\Sigma_0$ 
                                       \\ (5) \end{tabular} } &
\colhead{\begin{tabular}{c} $\Qsp$     \\ (6) \end{tabular} } &
\colhead{\begin{tabular}{c} $W/L_x$\tablenotemark{b} 
                                       \\ (7) \end{tabular} } &
\colhead{\begin{tabular}{c} $\lambda_{\rm y}/L_x$\tablenotemark{c} 
                                       \\ (8) \end{tabular} } &
\colhead{\begin{tabular}{c} $\lambda_{\rm y}/W$\tablenotemark{b,c} 
                                       \\ (9) \end{tabular} } &
\colhead{\begin{tabular}{c} $\Kymax$\tablenotemark{d}
                                       \\ (10)\end{tabular} }
}
\startdata
H1 & 1.7 & $\infty$ & 1 & 2.36 & 1.11 & 0.16 & 0.67 &  4.2 & 0.21 \\
H2 & 2.1 & $\infty$ & 2 & 3.91 & 1.06 & 0.06 & ...  &  ... & ...  \\
H3 & 2.3 & $\infty$ & 3 & 6.49 & 0.90 & 0.05 & ...  &  ... & ...  \\
    & & & & & & & & & \\
MS1 & 1.6 & 10      & 1 & 2.77 & 0.96 & 0.11 & 0.67 &  6.1 & 0.16 \\
MS2 & 1.8 & 10      & 2 & 4.91 & 0.81 & 0.07 & 0.22 &  3.1 & 0.31 \\
MS3 & 2.0 & 10      & 3 & 5.59 & 0.85 & 0.06 & 0.25 &  4.2 & 0.27 \\
    & & & & & & & & & \\
ME1 & 1.2 & 1       & 1 & 2.11 & 0.83 & 0.16 & 0.20 &  1.3 & 0.54 \\
ME2 & 1.4 & 1       & 2 & 2.45 & 0.89 & 0.16 & 0.22 &  1.4 & 0.49 \\
ME3 & 1.5 & 1       & 3 & 2.84 & 0.89 & 0.16 & 0.22 &  1.4 & 0.45 \\
ME4 & 1.6 & 1       & 3 & 2.72 & 0.97 & 0.16 & 0.25 &  1.6 & 0.45 \\
\enddata

\tablenotetext{a}{The prefixes H, MS, and ME stand for the
       hydrodynamic model and magnetized models with sub-equipartition
       ($\beta_0=10$) and equipartition ($\beta_0=1$)
       field strengths, respectively.}
\tablenotetext{b}{$W$ is the arm width at $\Sigma=\case{1}{2}
     (\Sigma_{\rm max}+\Sigma_{\rm min})$.}
\tablenotetext{c}{$\lambda_y$ is a mean separation along 
      the spiral arm of structures formed.}
\tablenotetext{d}{$\Kymax \equiv \ljsp/\lambda_y \rightarrow
   0.096 NQ_0 (\Smax/\Sigma_0)^{-1}$, where $N$ is the number of 
  structures formed within the simulation box 
  ($L_x=L_y/2 = \pi$ kpc).}

\end{deluxetable}

\clearpage
\begin{figure}
\epsscale{1.}
\caption{Schematic diagram showing the simulation domain in 
a typical two-armed spiral galaxy. The pitch angle $i$ between
the galactocentric circle and the spiral arm is assumed to be 
very small. The local rectangular box with a size $L_x\times L_y$
orbits with $\Omega_p = \Omega_0/2$ in the counterclockwise
direction. The arrows in the box indicate the linearly shearing velocity 
field due to the background
galactic differential rotation in the absence of the spiral
arm forcing. Our models adopt $R_0=10$ kpc and $\sin i=0.1$.
\label{arm_coord}}
\end{figure}

\begin{figure}
\epsscale{1.}
\caption{A sample equilibrium one-dimensional spiral shock profile 
(solid curves) for $Q_0=2.0$, $\beta_0=10$, and $F=3\%$. 
Dashed curve in (a) indicates 
the density profile for the non-self-gravitating counterpart.
Dotted curves in (a)-(e) represent the unperturbed (with $F=0$) solutions.
The dashed line in (b) indicates
the sound speed. See text for details.
\label{1Dshock10}}
\end{figure}

\begin{figure}
\epsscale{1.}
\caption{Same as Fig.\ \ref{1Dshock10} except
for $Q_0=1.5$, $\beta_0=1$, and $F=3\%$.
\label{1Dshock1}}
\end{figure}

\begin{figure}
\epsscale{1.}
\caption{Critical $Q_0$ values (solid lines and circles) and
$\Qsp\equiv Q_0 (\Sigma_{\rm max}/\Sigma_0)^{-1/2}$ 
(dotted lines and triangles) for marginally stable configurations.
Filled symbols indicate cases evidencing local dynamical instability,
while open symbols indicate incompatibility 
of equilibria with the adopted pattern speed ($\Omega_p=\Omega_0/2$).
\label{Qcrit}}
\end{figure}

\begin{figure}
\epsscale{1.}
\caption{(a) Initial surface density distributions and (b) evolution of
maximum surface density for hydrodynamic simulations.
\label{HDevol}}
\end{figure}

\begin{figure}
\epsscale{1.}
\caption{Snapshots of model H1 at $t/t_{\rm orb}=4.7$ ({\it left})
and model H3 at $t/t_{\rm orb}=4.0$ ({\it right}).
Surface density $\Sigma/\Sigma_0$ is shown in logarithmic scale.
Model H1, with a smaller $Q_0$ and a smaller arm-to-interarm 
density contrast, forms three weak spurs that grow downstream 
in the direction perpendicular to the spiral arm, while
model H3, having a higher $Q_0$ and a larger density contrast,
remains stable. The vertical, dotted line in model H1 marks
the peak density in the initial equilibrium,
and solid lines overlaid on the spur wavefronts are from equation 
(\ref{Wfront}) with $\kx(0)/\ky=-2.2.$
\label{HDimage}}
\end{figure}

\begin{figure}
\epsscale{1.}
\caption{(a) Initial surface density distributions and (b) evolution of
maximum surface density for MHD simulations with $\beta_0=10$.
\label{MHD10evol}}
\end{figure}

\begin{figure}
\epsscale{1.}
\caption{ A snapshot of model MS3 at $t/\torb = 1.78$.
(a) Surface density and velocity 
fields seen in the frame comoving with the spiral arm pattern.
Streaming motions and reversed shear due to the arm, as well as
normal background shear outside the arm are apparent in
the velocity field.
(b) A few selected streamlines (dotted lines)
in the frame comoving with spurs, together with
wavefronts (solid lines) as defined by equation (\ref{Wfront}) 
with $\kx(0)/\ky=2.5$, are overlaid
on surface density;
the left boundary is shifted to correspond to the shock location.
In both panels, surface density is displayed in logarithmic color scale.
\label{MHD10pat}}
\end{figure}

\begin{figure}
\epsscale{1.}
\caption{Final density structures of MHD simulations with $\beta_0=10$
for models MS1 at $t/\torb = 4.0$ ({\it left})
and MS2 at $t/\torb = 2.1$ ({\it right}). Magnetic
field lines are drawn in red lines and scalebars label
$\log{(\Sigma/\Sigma_0)}$.
\label{MHD10image}}
\end{figure}

\begin{figure}
\epsscale{1.}
\caption{(a) Initial surface density distributions and (b) evolution of
maximum surface density for MHD simulations with $\beta_0=1$.
\label{MHD1evol}}
\end{figure}

\begin{figure}
\epsscale{1.1}
\caption{A snapshot of model ME3 at $t/\torb = 1.27$.
(a) Density structure in logarithmic color scale and
velocity fields (vectors) in the spiral arm frame. 
(b) Wavefronts of spurs (solid lines; eq.\ [\ref{Wfront}] with $\kx(0)/\ky=0.2$)
and selected streamlines (dotted lines) in the frame
comoving with spurs. 
(c) A synthetic color map of the line-of-sight velocity in the units of 
km s$^{-1}$, overlaid on surface density contours spaced at 
$\Sigma=2.5, 5, 7.5, 10, 12.5\Msun\pc^{-2}$; 
the viewing direction is parallel to the arm. 
(d) Same as (c) except
that the viewing angle is oblique to the arm. See text for details.
\label{MHD1pat}}
\end{figure}

\begin{figure}
\epsscale{1.}
\caption{Amplification factors of the MJIs within a spiral arm
are drawn 
with solid contours, spaced at 
100, 90, ..., 10, from inside to outside.
Dotted contours show corresponding growth times 
$\tau_{\rm grow}=1.0$, 1.5, ..., 5.0, from right to left.
The chosen condition for background flow is the same as in model ME3.
A thick box near $K_y=0.45$ and $\kx(0)/\ky=0.2$ marks the 
most unstable mode apparent in the nonlinear model ME3,
with the length of its each side representing an uncertainty in
determining $K_y$ and $\kx(0)/\ky$ from the simulation outcome.
\label{mhd_con}}
\end{figure}

\begin{figure}
\epsscale{1.}
\hspace{9cm}
\caption{Comparison of the linear analysis with the simulation results of 
model ME3 at $t/\torb=1.27$. 
Linear evolution of the perturbed density (solid line; 
$K_y=0.45$ and $\kx(0)/\ky=0.2$) is drawn 
as a function of the dimensionless
Lagrangian time variable $\tau$ (lower $x$-axis)
or the distance from the shock front (upper $x$-axis). 
Various curves (dotted, dashed, dot-dashed) plot the perturbed 
surface density profiles, with an offset of 1.2, along three prominent spurs
shown in Figure \ref{MHD1pat}.
\label{bc_comp}}
\end{figure}

\begin{figure}
\epsscale{1.}
\caption{Final density structures of MHD simulations with $\beta_0=1$
for models ME3 at $t/\torb = 1.59$ ({\it left})
and ME4 at $t/\torb = 2.55$ ({\it right}). Magnetic
field lines are drawn in solid lines and colorbars label
$\log{(\Sigma/\Sigma_0)}$.
\label{MHD1image}}
\end{figure}

\end{document}